\documentclass[aps,prl,twocolumn,superscriptaddress]{revtex4-1}

\usepackage{graphicx}
\usepackage{amsmath}
\usepackage{array}
\usepackage{color}

\begin{document}

\title{Surface enhanced circular dichroism spectroscopy mediated by non-chiral nanoantennas}

\author{Aitzol Garc\'{i}a-Etxarri}
\affiliation{Department of Materials Science and Engineering, Stanford University, Stanford, California 94305, United States}
\email[]{aitzol@stanford.edu}

\author{Jennifer A. Dionne}
\affiliation{Department of Materials Science and Engineering, Stanford University, Stanford, California 94305, United States}

\date{\today}

\begin{abstract}

We theoretically investigate light-matter interactions for chiral molecules in
the presence of non-chiral nanoantennas. Isotropic nanostructures
supporting optical-frequency electric or magnetic dipoles are
sufficient to locally enhance the excitation of a molecule's chiral
polarizability and thus its circular dichroism spectrum. However,
simultaneous electric and magnetic dipoles are necessary to achieve
a net, spatially-averaged enhancement. Our contribution provides a
theoretical framework to understand chiral light-matter
interactions at the nanoscale and sets the necessary and sufficient
conditions to enhance circular dichroism spectroscopy in the
presence of nanoantennas. The results may lead to new,
field-enhanced, chiral spectroscopic techniques.

\end{abstract}


\maketitle

Chiral objects - those whose mirror images are not superimposable -
abound in nature. Examples range from spiral galaxies to human
hands to nucleic acids. Chirality is fundamental to many
biological, chemical, and physical processes: it can contribute to
the gene flow and evolution of snail colonies
\cite{Ueshima:2003bc}, determine the effectiveness of
pharmaceutical drugs, and impact the formation of fermionic
condensates and superfluids. Nature's preference for a certain
chirality, e.g. the fact that all proteins form a left-handed
spiral while sugars twist to the right, constitutes one of life's
greatest mysteries \cite{Hegstrom:1990}.

In electromagnetism, circularly-polarized light (CPL) is the paradigmatic
example of a chiral field. Propagating in a dispersion-less medium,
the electric and magnetic fields of CPL undergo one full rotation
per period and wavelength. This rotation can be either clockwise or
counter-clockwise. The spatial trajectories of these two waves form a
chiral set of solutions to Maxwell's equations.

Circularly polarized light  can be used to probe the geometric and
electromagnetic chiral properties of molecules. A specific
enantiomer of a chiral molecule will exhibit a preferential
absorption of right or left-handed circularly polarized light.
Circular dichroism (CD) spectroscopy measures this differential
absorption in the ultraviolet and visible spectrum, while
vibrational circular dichroism (VCD) extends the technique to the
infrared \cite{Barron:2004vd}. Both CD and VCD are highly valuable
techniques: for example, in molecular biology, they elucidate a protein's
secondary structure, which in turn provides insight into the
protein's function. In pharmacology, these techniques determine the chiral
purity of chemical products and the absolute structure of pure
enantiomers \cite{Rodger:2006jo, Polavarapu:2000tf}.

Despite the wide applicability of CD and VCD, their sensitivity is
 limited. Proteins and nucleic acids exhibit a differential absorption
of left and right circularly polarized light
which is nearly five orders of magnitude less
than their absorption of unpolarized light. Hence, in order to get measurable signals, large sample concentrations or
signal amplifiers are needed. This prohibits CD and VCD
spectroscopy from entering the few molecule regime.

In recent years, engineered light-matter interactions at the
nanoscale have enhanced the sensitivity of other spectroscopic
techniques such as Surface Enhanced Raman Spectroscopy (SERS) and
Surface-Enhanced Infra-Red Absorption spectroscopy (SEIRA).
Nanostructured surfaces and nanoparticles supporting strong
optical-frequency electric resonances have allowed SERS and SEIRA
to reach single molecule and attomolar sensitivity, respectively
\cite{Nie:1997ub, Kneipp:1997bs, Xu:1999wr, Neubrech:2008co}. In this paper, drawing on insights from SERS and SEIRA,
we investigate the fundamental electromagnetic conditions
necessary to increase the sensitivity of CD and VCD spectroscopy.
Since chiral molecules are characterized by electric and magnetic
dipoles, enhancing (V)CD spectroscopy requires control over both the electric
\emph{and} magnetic fields of light. Accordingly, attention is 
given to nanostructures supporting either
electric \emph{or} magnetic resonances, or both. We show that molecules
near an electrically- or magnetically-resonant nanoparticle will
experience a local, position-dependent chiral enhancement. Averaged
over all space, such electrically- or magnetically-resonant
nanoparticles do not produce a net enhancement of the (V)CD
signal. In contrast, engineered nanostructures with combined electric and magnetic resonances
do produce a global enhancement of the signal. Our
results determine the necessary and sufficient conditions required to
enhance chiral light-matter interactions, and may enable new
spectroscopic techniques such as field enhanced circular dichroism
 or field enhanced vibrational circular dichroism.

To describe the excitation of chiral molecules by electric and
magnetic fields of arbitrary polarization states, we adopt the
formalism introduced by Tang and Cohen \cite{Tang:2010cw, Tang:2011df}.
 In general, we assume that a circularly polarized  plane wave
illuminates a nanoparticle, and that fields of arbitrary
polarization will be generated in the surroundings of the particle. The rate of
excitation of a randomly oriented chiral molecule illuminated by
these fields can be expressed as:
\begin{equation}\label{A}
A^\pm=\frac{1}{\varepsilon_0}\omega U_e^{\pm}\alpha''+\mu_0\omega U_m^{\pm} \chi''+\frac{2}{\varepsilon_0}C^{\pm}G'' 
 \end{equation}
where the plus and minus superscripts denote whether the fields
originated from left- or right-handed circularly polarized incident
light, respectively.
$U_e^{\pm}=\frac{\varepsilon_0}{2}|\mathbf{E^{\pm}}|^2$ and
$U_m^{\pm}=\frac{1}{2\mu_0}|\mathbf{B^{\pm}}|^2$ are the electric
and magnetic energy densities respectively, $\alpha''$ is the
imaginary part of the electric polarizability, $\chi''$ is the
imaginary part of the magnetic susceptibility, and $G''$ is the
imaginary part of the isotropic mixed electric-magnetic dipole
(i.e., chiral) polarizability. As usual, $\omega$, $\varepsilon_0$,
and $\mu_0$ denote the angular frequency of light and the
permittivity and permeability of free space. $C^{\pm}$ denotes the
electromagnetic density of chirality and can be defined as
\cite{Tang:2010cw, Bliokh:2011vi}:


\begin{equation}\label{C}
C^{\pm}=\frac{1}{2c^2}\omega Im(\mathbf E^{*\pm} \cdot \mathbf H^{\pm})=\frac{\varepsilon_0}{2}\omega |\mathbf E^{\pm}| |\mathbf H^{\pm}|cos(\beta_{i\mathbf{E^{\pm}},\mathbf{H^{\pm}}})
 \end{equation}

In the above expression, $\beta_{i\mathbf{E},\mathbf{H}}$ denotes
the angle between the complex $\mathbf{E}$ vector multiplied by the
complex number $i$ and the $\mathbf{H}$ vector. 

The three terms of Eq.~\ref{A} account for molecular absorption due
to electronic excitations of electric, magnetic, and chiral
character respectively. For most molecules, $\chi$ is negligible,
and the second term in Eq.\ref{A} can be neglected.
In this paper, we focus on techniques to enhance the third term,
proportional to $C$, corresponding to chiral light-molecule
interactions.

For circularly polarized light traveling in vacuum,
$C_{cpl}=\pm\frac{\varepsilon_0 \omega}{c}E^2_0$ where $E_0$ is the
incoming electric field amplitude. Such propagating solutions of
Maxwell's equations satisfy
$cos(\beta_{i\mathbf{E},\mathbf{H}})=\pm1$, indicating, as recently
pointed out by others \cite{Andrews:2012ks, Bliokh:2011vi,
Hendry:2012kv}, that CPL is an optimum polarization state to
maximize the excitation rate of chiral transitions in chiral
molecules. However, Eq.\ref{C} shows that it is possible to enhance
chiral light-matter interactions by enhancing the electric and/or
magnetic fields, provided their polarization state and relative
phase remain approximatively unaltered.

Since CD measures the differential absorption of a
system excited by right and left-handed CPL, the measured signal can be expressed as
\begin{align}\label{CD_gen}
CD\propto A^+ - A^-=\frac{\omega}{\varepsilon_0}\alpha''(U_e^{+}- U_e^{-})+\frac{2}{\varepsilon_0}G''(C^+-C^-)
\end{align}
$C$ is an electromagnetic pseudo scalar (flips sign under parity),
while $U_e$ instead is a pure scalar (does not flip the sign)
\cite{Tang:2010cw, Tang:2011df}. Accordingly, for circularly polarized light,
$CD_{cpl}=~\frac{4}{\varepsilon_0}G''|C_{cpl}|$. As seen, the
measured signal is proportional to the chiral polarizability of the
molecule ($G$), allowing CD techniques to unveil the chiral
electromagnetic properties of molecules.

In this letter, we investigate this process in the presence of
isotropic, non-chiral nanoantennas. Due to the isotropic nature of
the antenna,
 a parity inversion on the system composed of the
particle and incident fields will exclusively change the handedness
of the incident fields. Consequently, exciting the antenna with
left and right CPL will be equivalent to applying a parity
operation over the system. Finally, due to the symmetry properties
of $C$ and $U_e$, the electromagnetic fields in the presence of an
achiral antenna at an arbitrary position will hold the following properties:  $C^+ = C =
-C^-$ and $U^+_e= U^-_e$, and  Eq.\ref{CD_gen} can be simplified to
\begin{align}\label{CD}
CD\propto A^+ - A^-=\frac{4}{\varepsilon_0}G''C.
\end{align}

Eq.\ref{CD} can be rewritten as $CD=f CD_{cpl}$, where
$f=\frac{C}{|C_{cpl}|}$. In the presence of an antenna, this is a
position dependent quantity. Hence, from now on we will refer to
this quantity as the local CD enhancement factor $f(\mathbf{r})$.
Note that if the antenna were by any means chiral
\cite{Plum:2009ks}, the the parity relations for $C$ and $U_e$
would not be fulfilled and Eq.\ref{CD} would not be valid to
describe CD spectroscopy in the presence of a particular
nanoantenna.

Can a nanoparticle enhance CD spectroscopy? Recent circular dichroism
studies performed on molecules in the presence of chiral \cite{Hendry:2012kv, Hendry:2010gy, Klimov:2012vt, Guzatov:2012wp, Schaferling:2012bs} and
plasmonic nanostructures \cite{Slocik:2011ei, Zhang:2012uc, Govorov:2010fb, Govorov:2011wj, Govorov:2012kp, Lieberman:2008df} suggest that the approach has potential. However, a fundamental understanding of these processes is
still lacking.

Let us start by considering the optical response of a small
plasmonic nanoparticle. If the size of the nanoparticle is much
smaller than the wavelength of light, the electrostatic
approximation holds and the response of the sphere can be modeled
as an electric dipole. Assuming that the particle is illuminated by
a plane wave of arbitrary polarization,
 the sphere will acquire an electric dipolar
moment $\mathbf p_e=\alpha_e\mathbf {E_{inc}}$, where $\mathbf
{E_{inc}}$ is the incoming electric field vector,
$\alpha_e=4\pi\varepsilon_0 a^3 \frac{\varepsilon
-1}{\varepsilon+2}$ is the electric polarizability of the sphere in
vacuum, $\varepsilon$ is the permittivity of the particle, and $a$
is the radius of the sphere.

In the near-field of the electrically-resonant nanoparticle, the
total fields can be expressed as the sum of scattered and incoming
fields. Since the magnetic near fields scattered by an electric dipole
are negligible in the electrostatic limit, we find:
\begin{equation}\label{Etot_ED}
\mathbf{E_{tot}}=\mathbf{E_{scat}}+\mathbf{E_{inc}}=\frac{k^2}{\varepsilon_0}\mathbf{G_ep_e}+\mathbf{E_{inc}}
\end{equation}
\begin{equation}\label{Htot_MD}
\mathbf{H_{tot}}=\mathbf{H_{scat}}+\mathbf{H_{inc}}\approx \mathbf{H_{inc}},
\end{equation}
where $k$ is the wavenumber in vacuum and $\mathbf{E_{scat}}$,
$\mathbf{H_{scat}}$, and $\mathbf{H_{inc}}$ are the scattered and
incident electric and magnetic field vectors. $\mathbf{G_e}$ is the
electric dyadic Green's tensor in the near field limit, given by \cite{Ge}.
For simplicity, we assume that the sphere is located at the origin
of coordinates ($\mathbf{r_0}=0$) and is illuminated by a plane wave
traveling in the positive $\hat{z}$ direction.

\begin{table*}
\footnotesize
   \begin{tabular}{| c | c | }
    \hline
     & $f(\mathbf{r})=\frac{C(\mathbf{r})}{C_{cpl}}$ \\ \hline
    $\alpha_e$ & $f^{\alpha_e}(\mathbf{r})=\pm\left[1+\frac{1}{2}\frac{1}{4\pi\varepsilon_0}\frac{1}{r^3}(1-3cos^2\theta)|\alpha_e| cos(\varphi_e)\right]$ \\ \hline
    $\alpha_m$  & $f^{\alpha_m}(\mathbf{r})=\pm\left[1+\frac{1}{2}\frac{1}{4\pi}\frac{1}{r^3}(1-3cos^2\theta) |\alpha_m| cos( \varphi_m)\right]$ \\ \hline
    $\alpha_e+\alpha_m $ & $f^{\alpha_e+\alpha_m}(\mathbf{r})=f^{\alpha_e}(\mathbf{r})+f^{\alpha_m}(\mathbf{r}) \pm \left[ -1+ \frac{1}{2}\frac{1}{4\pi\varepsilon_0}\frac{1}{4\pi}\frac{1}{r^6}(5-3cos^2\theta)|\alpha_e| |\alpha_m| cos( \varphi_m-\varphi_e) \right]$ \\ \hline
    \end{tabular}   \caption{Analytic expressions for the position dependent CD enhancement factor $f(\mathbf{r})$ around different basic electromagnetic modes. $\alpha_e$ and $\alpha_m$ account for an electric and a magnetic polarizability respectively. $\varphi_e$ and $\varphi_m$ represent the phase of $\alpha_e$ and $\alpha_m$ respectively. $\theta$ is the spatial elevation angle (see inset in Fig.\ref{Ag_CPL}a).} \label{C_r_table}
\end{table*}
For circularly polarized illumination, the local density of
chirality $C$ in the near field of the sphere can be analytically
calculated by substituting Eqs.\ref{Etot_ED} and \ref{Htot_MD} into
Eq.\ref{C}. The resulting expression for the position dependent CD
enhancement is given in Table \ref{C_r_table} as
$f^{\alpha_e}(\mathbf{r})$.

\begin{figure}
\centering
 \includegraphics[width=1\columnwidth]{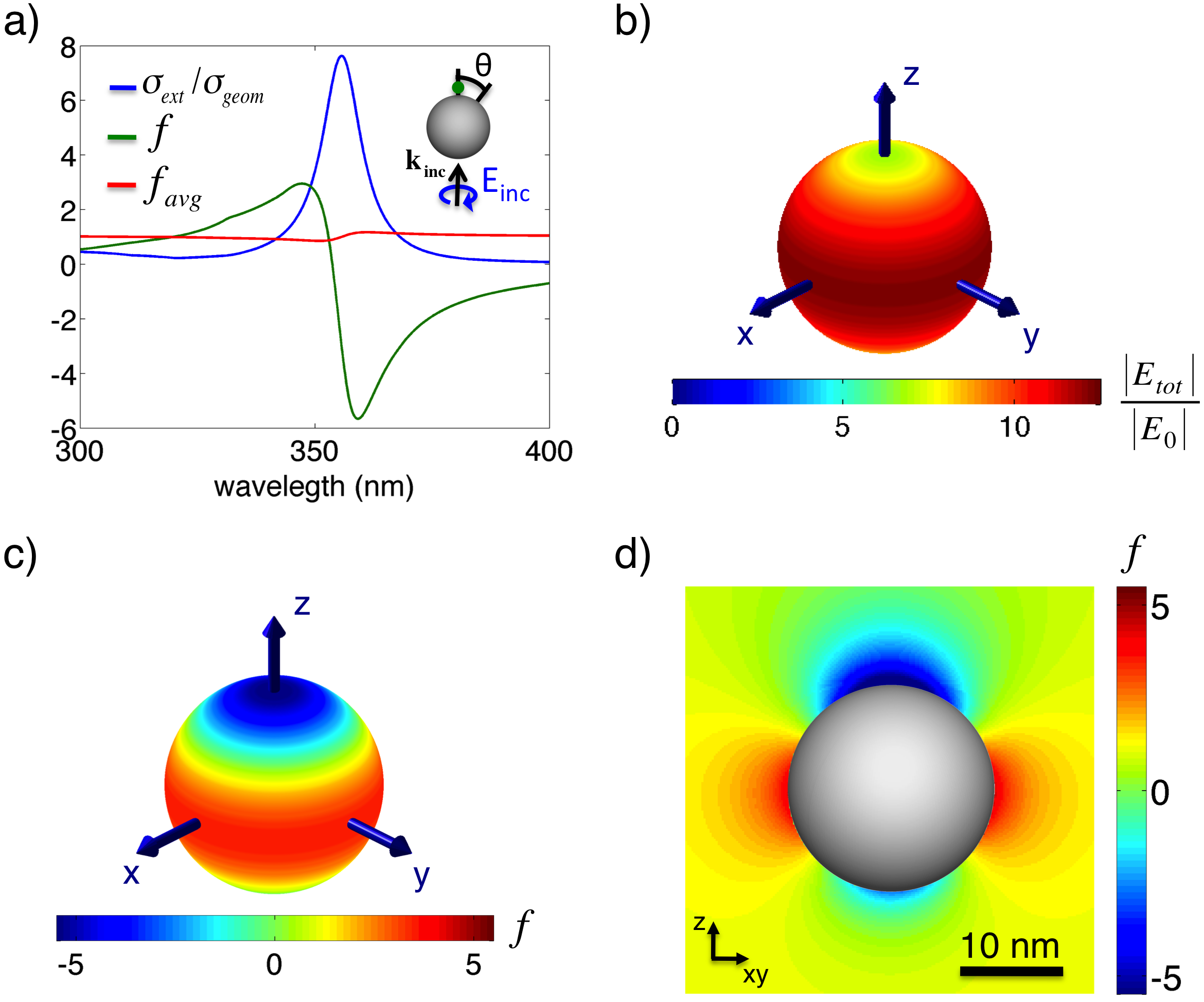}
 \caption{\label{Ag_CPL} 10 nm radius silver nano
sphere illuminated by a circularly polarized plane wave. a) Spectral
response of the nanostructure. Blue: geometrically normalized
extinction cross section. Green: CD enhancement
factor, $f(\mathbf{r})$, $1$ nm on top of the sphere (green dot in the inset). Red: CD
enhancement factor averaged over a sphere covering the particle ($f_{avg}$). b)
Total electric near field amplitude enhancement at the plasmonic resonance
($\lambda=359$ nm) plotted over a surface $1$  nm above the
nanoparticle. c) CD enhancement factor, $f(\mathbf{r})$, at that same
surface and wavelength. d) $f(\mathbf{r})$ for the same wavelength plotted over a plane crossing the
sphere and the incident wavevector.}
 \end{figure}

As an example, consider the response of a $10$ nm radius silver
sphere, determined by rigorously solving Maxwell's equations using
the Boundary Element Method (BEM) \cite{deAbajo:2002ud,
deAbajo:1998um}. Fig.\ref{Ag_CPL}a plots the extinction cross
section of the sphere, as well as the position-dependent CD
enhancement factor, $f(\mathbf{r})$, calculated at a point $1$ nm above the particle, and
the average enhancement, defined as the local enhancement
integrated over a surface, normalized to the surface area:
$f_{avg}=\int f(\mathbf{r})dS/(4\pi r^2)$. As seen, the localized surface
plasmon resonance occurs at a wavelength of $\lambda=355$ nm, where
the extinction cross-section is maximum. Here, the absolute value
of the density of chirality is minimized. Since the electric dipole
is on-resonance, the phase of $\mathbf{E_{scat}}$ will be $\pi/2$
delayed with respect to the incoming magnetic field. Therefore,
the phase of the total (incident and scattered) fields results in
$cos(\beta_{i\mathbf{E},\mathbf{H}})\approx0$ and, consequently,
according to Eq.\ref{C},
$f\propto C\approx0$. 

Slightly off the plasmonic resonance, however, the ideal phase
relation between the electric and magnetic field is not completely
destroyed, and a local enhancement of $f(\mathbf{r})$ can be obtained.
Fig.\ref{Ag_CPL}b and c show the total electric field enhancement and the
CD enhancement factor $f(\mathbf{r})$, plotted along a spherical surface $1$  nm
away from the particle for $\lambda=359$ nm, red-shifted slightly
from the plasmon resonance of $\lambda=355$ nm. As seen, $f(\mathbf{r})$  is both positively and negatively enhanced in
different regions around the nanoparticle. Here, a negative
enhancement of $f(\mathbf{r})$ indicates that a left-handed excitation will
induce right-handed near fields. Positive enhanced values of $f(\mathbf{r})$
imply that the chirality will be increased while the
handedness of the fields will remain unchanged.

Neglecting retardation of the incident electromagnetic fields in
the vicinity of the particle, the analytic average density of
CD enhancement is found to be $f_{avg}=1$. Fig.\ref{Ag_CPL}a (red
line) plots this averaged CD enhancement factor for the fully
electrodynamically calculated fields around the particle. The small
deviations from the predicted analytical value of $1$ are due to
retardation effects.
 Therefore, a plasmonic nanoparticle illuminated by circularly
polarized light will not enhance CD spectroscopy if isotropically
surrounded by chiral molecules. However, surface-enhanced CD
spectroscopy could be achieved by creating a surface of
closely-packed nanoparticles, where molecules are
prohibited from laying in regions of opposite $f(\mathbf{r})$. The specifics of the
design of such a surface will be left for future work. 

 Now consider an ideal optical-frequency
magnetic dipole illuminated by CPL. Expressions for the CD enhancement factor are included in Table \ref{C_r_table}
as $f^{\alpha_m}(\mathbf{r})$. 
As with electric dipoles, local
enhancement of $f(\mathbf{r})$ is possible, but averaging the CD enhancement
factor on a sphere of constant radius yields no net enhancement.

This situation is significantly modified when considering a
combination of an isotropic electric and an isotropic magnetic
dipole. If the dipoles are located at the same spatial position,
symmetry dictates that they will not interact; accordingly
bi-anisotropy will not be generated \cite{sheikoleslami:2011}. In
the electrostatic limit, assuming that the magnetic fields produced
by an electric dipole as well as the electric fields produced by a
magnetic dipole are negligible, the total near fields produced by
such a system can be approximated as:
\begin{equation}\label{Etot_EDMD}
\mathbf{E_{tot}}=\mathbf{E_{scat}}+\mathbf{E_{inc}}\approx \frac{k^2}{\varepsilon_0}\mathbf{G_ep_e}+\mathbf{E_{inc}}
\end{equation}
\begin{equation}\label{Htot_EDMD}
\mathbf{H_{tot}}=\mathbf{H_{scat}}+\mathbf{H_{inc}}\approx k^2\mathbf{G_ep_m}+\mathbf{H_{inc}},
\end{equation}

For circularly polarized incident light, averaging the position
dependent $f(\mathbf{r})$ ($f^{\alpha_e+\alpha_m}(\mathbf{r})$ in
Table \ref{C_r_table}) gives the following expression:
\begin{align}\label{C_EDMD_int}
f^{\alpha_e+\alpha_m}_{avg}= \left[ 1+ \frac{1}{4\pi\varepsilon_0} \frac{1}{2\pi} \frac{1}{r^6} |\alpha_e||\alpha_m|cos\left( \varphi_m-\varphi_e\right) \right]
\end{align}
which in general can be greater than $1$. $\varphi_m$ and $\varphi_e$ account for the phase of the magnetic and electric polarizability respectively.

 \begin{figure}
\centering
 \includegraphics[width=1\columnwidth]{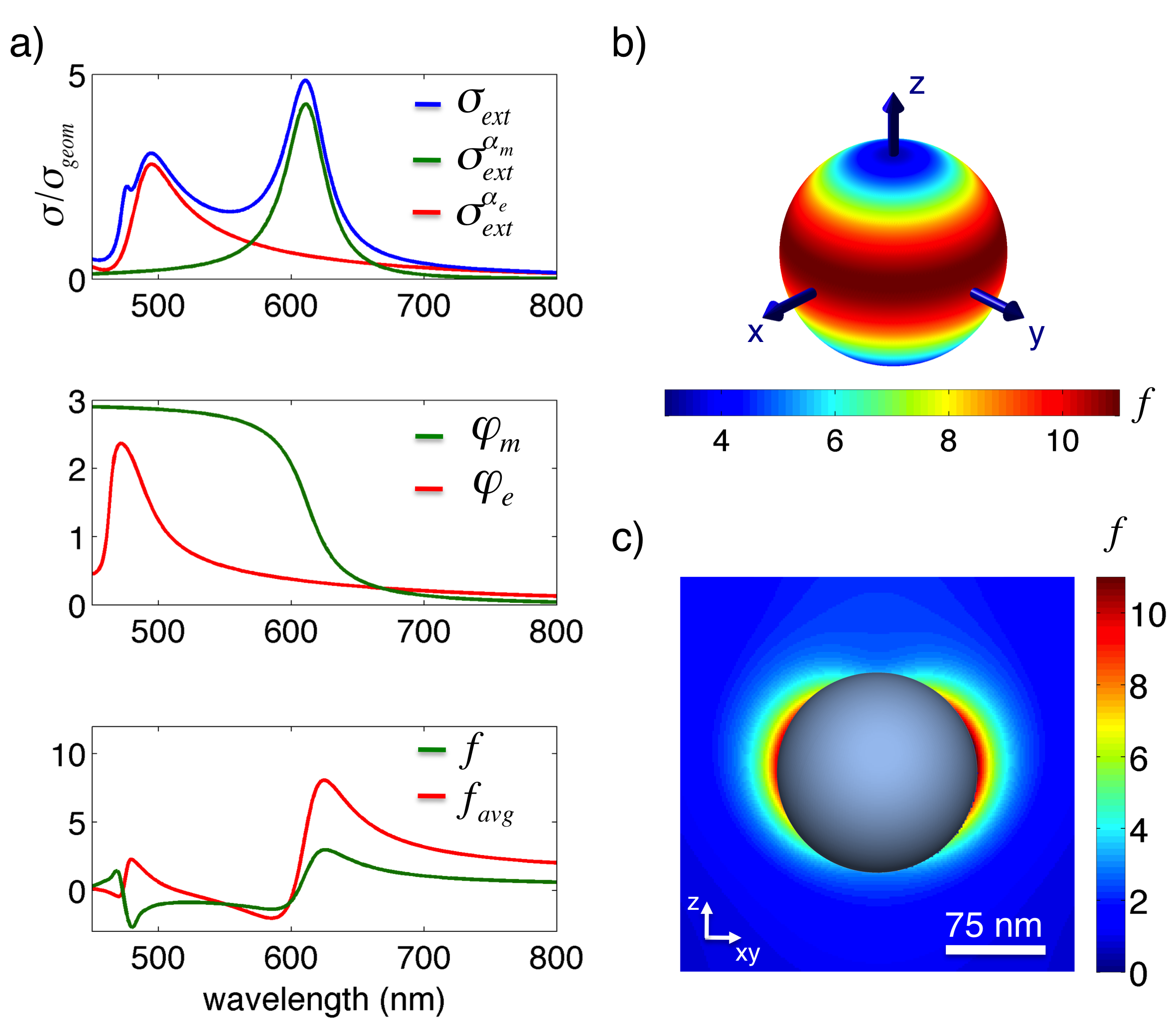}
 \caption{\label{Si_CPL} $75$ nm radius silicon nanosphere illuminated by circularly polarized light. a) Upper panel: The geometrically normalized extinction cross section is plotted in blue. Green and red lines plot the magnetic and electric dipolar contributions to the total extinction cross section respectively. Middle panel: phase behavior of the magnetic (green) and electric (red) dipolar contributions. Lower panel: CD enhancement factor $f(\mathbf{r})$ at a position $1$nm on top of the sphere 
 and averaged CD enhancement $f_{avg}$ plotted in green and red respectively. The spatially averaged CD enhancement factor is enhanced in the spectral regions where $cos(\varphi_m-\varphi_e) \sim \pm 1$.  b) CD enhancement factor at a surface covering the nanoparticle for a separation distance of $1$ nm at $\lambda=625 nm$. c) CD enhancement factor for the same wavelength plotted over a plane crossing the sphere. }
 \end{figure}
While optical-frequency magnetic dipoles are uncommon in nature,
recently-engineered nanostructures can exhibit strong electric and
magnetic resonances at visible and near-infrared frequencies.
Examples include split-ring resonators, nanocups or
nanocrescents \cite{Zhang:2011bo, Shevchenko:2009}, as well as
rings of metallic nanoparticles \cite{Manoharan:2010us}.
As an example of a system showing isotropic electric and magnetic
resonances, we consider a spherical silicon nanoparticle
\cite{GarciaEtxarri:2011bo, Evlyukhin:2012, Kuznetsov:2012}. Due to
their high refractive index, these particles support electric and
magnetic Mie resonances in the visible and infrared part of the
spectrum \cite{GarciaEtxarri:2011bo}. The upper panel of
Fig.\ref{Si_CPL}a shows the geometrically normalized extinction
cross section of a $75$ nm radius Si nanoparticle, along with the
contribution of the lowest order electric ($\alpha_e$) and magnetic
($\alpha_m$) modes to the total extinction spectrum, calculated
through Mie theory \cite{Bohren:2008wi}. The middle panel
illustrates the phase behavior of the polarizabilities $\alpha_e$
and $\alpha_m$. According to Eq.\ref{C_EDMD_int}, the averaged
chiral density will be increased only if $cos(\varphi_e-\varphi_m)
\sim \pm1$ while $|\alpha_e||\alpha_m|>1$, a condition which is
satisfied in the green and red regions of the visible spectrum. The
lower panel plots the local chiral density at a point 1 nm above
the nanoparticle (green) and the averaged chiral density (red)
around the particle. As seen, unlike an isolated electric or
magnetic dipole, the total averaged chiral density is enhanced
compared to CPL.

Fig.\ref{Si_CPL}b and c plot the three-dimensional surface ($1$~
nm above the sphere) and two-dimensional cross-section of $f(\mathbf{r})$ for
$\lambda=625$ nm. At this wavelength, the maximum local and
averaged CD enhancement factor occur. Remarkably, $f(\mathbf{r})$ is
exclusively positive throughout all space at this wavelength.
Moreover, the averaged enhancement factor exceeds that of
circularly polarized light by nearly an order of magnitude. In
other words, solution-based CD spectroscopy can be
enhanced globally via isotropic achiral nanostructures supporting
electric and magnetic dipoles.

In conclusion, we have shown how individual electric or magnetic
dipoles can tailor the local density of chirality and thus the
circular dichroism enhancement. Individual electric or magnetic
dipoles can only locally (but not globally) increase CD enhancement
factor. Consequently, they could be used to increase the
sensitivity of surface-based circular dichroism techniques, but are
ill-suited for solution-phase measurements. In contrast,
combinations of electric and magnetic dipoles can achieve both
local and net, global enhancements of the CD signal, and are
therefore strong candidates for solution-phase surface enhanced
(V)CD. Si nanoparticles, for example, increase the averaged CD
enhancement factor by nearly a factor of ten. Further engineering
of the electric and magnetic response of nanoantennas may lead to
much higher enhancement factors. Interestingly, we demonstrate that
structurally chiral particles are not necessary to enhance CD
spectroscopy. These results set the theoretical basis for
field-enhanced chiral spectroscopic techniques, providing a new
twist to molecular probes in fields ranging from molecular biology
to pharmacology.


  \begin{acknowledgements}
We thank Sassan Sheikholeslami and Hadiseh Alaeian for insightful discussions and the Donostia International Physics Center (DIPC) for their computational resources. Funding from a NSF Career Award (DMR-1151231) and Stanford's Global Climate and Energy Project are gratefully acknowledged. 
  \end{acknowledgements}

 \bibliography{Enhancing_CD}

\begin{thebibliography}{36}
\bibitem [{\citenamefont
  {$\mathbf{G_{e}p}=\frac{1}{4\pi}\frac{e^{ikr}}{k^2r^3}\left(3\mathbf{n}(\mathbf{n}\cdot~\mathbf{p})
  \mathbf{p}\right)$}()}]{Ge}%
  \BibitemOpen
  \bibfield  {author} {\bibinfo {author} {\bibnamefont
  {$\mathbf{G_{e}p}=\frac{1}{4\pi}\frac{e^{ikr}}{k^2r^3}\left(3\mathbf{n}(\mathbf{n}\cdot~\mathbf{p})
  \mathbf{p}\right)$}},\ }\href@noop {} {\bibinfo  {journal} {where
  $\mathbf{p}$ represents a generic dipolar moment (electric or magnetic),
  $r=|\mathbf{r}-\mathbf{r_0}|$ is the distance between the position of the
  dipole ($\mathbf{r_0}$) and the observation point $\mathbf{r}$, and
  $\mathbf{n}$ is the unit vector in the direction of
  $|\mathbf{r}-\mathbf{r_0}|$}\ }\BibitemShut {NoStop}%
\end{thebibliography}

\end{document}